\documentclass[twocolumn,aps,prl,floatfix,showpacs]{revtex4}
\usepackage{graphicx,graphics,color,epsfig}
\usepackage{amsmath}
\usepackage{amssymb}

\begin{document}
\def\runtitle{MC algorithm in quasi-one-dimensional systems}
\def\runauthor{Tota Nakamura}

\title{%
Efficient Monte Carlo algorithm in quasi-one-dimensional Ising spin systems
}

\author{Tota Nakamura}

\affiliation{%
Faculty of Engineering, Shibaura Institute of Technology,
         Minuma-ku, Saitama  330-8570, Japan
}

\date{\today}

\begin{abstract}
We have developed an efficient Monte Carlo algorithm, which accelerates 
slow Monte Carlo dynamics in quasi-one-dimensional Ising spin systems.
The loop algorithm of the quantum Monte Carlo method is applied to
the classical spin models with highly anisotropic exchange interactions.
Both correlation time and real CPU time are reduced drastically.
The algorithm is demonstrated 
in the layered triangular-lattice antiferromagnetic Ising model.
We have obtained
the relation between the transition temperature and the 
exchange interaction parameters, which modifies the result of 
the chain-mean-field theory.
\end{abstract}

\pacs{05.10.Ln, 05.50.+q, 75.40.Mg. }
\maketitle

The application of the Monte Carlo (MC) method to the condensed-matter physics
has been successful bridging between the experimental study and the
theoretical study.\cite{binder}
The simulational results are now quantitatively 
compared with the experimental results.
We may estimate various physical parameters,
predict unknown properties, and propose new experiments on real materials.
However,
we encounter a difficulty when we apply the MC method to the frustrated 
systems.
The MC dynamics slows down, and 
it becomes very hard to reach the equilibrium states.
Since frustration has been recognized to play an important role in 
novel effects of many materials,\cite{HFM2006}
we somehow have to overcome this difficulty to study
new properties, new concepts and new function of such materials.

In this Letter we consider the quasi-one-dimensional (Q1D) 
frustrated spin systems.
The magnetic exchange interaction of this system is highly anisotropic.
The interaction along the $c$ axis is much stronger than those within
the $ab$ plane: $|J_c| \gg |J_{ab}|$.
The experimental realizations of this model are the
ABX$_3$-type compounds.\cite{CsCoBr3,CsCoCl3,visser,experi-mag,experi-f}
The lattice structure is the stacked triangular lattice with the
antiferromagnetic exchange interactions.
There are two reasons for the slow MC dynamics in this system.
One is frustration, and the other is the long correlation length along the
$c$ axis.
The single-spin-flip algorithm cannot change the states of these
correlated clusters.
Koseki and Matsubara \cite{chb1,chb2,koseki} 
proposed the cluster-heat-bath method,
which accelerates the MC dynamics in Q1D Ising spin
systems.
When we update a spin state, 
the transfer matrix is multiplied along the $c$ axis.
This matrix operation takes a long CPU time.
The possible size of simulation has been restricted to
the system with
$|J_c/J_{ab}|=10$, $36\times 36\times 360$ spins, 
and $2\times 10^6$ MC steps.\cite{meloche}   
Considering that the ratio $|J_c/J_{ab}|$ in real compounds is
in the order of 100, we need to develop another algorithm that
improves the simulation efficiency.

We notice that the similar slow-dynamic situation occurs in the
quantum Monte Carlo (QMC) simulation.\cite{suzukitrotter}
The $d$-dimensional quantum system is mapped to the $(d+1)$-dimensional
classical system, on which the simulation is performed.
The additional dimension is called the Trotter direction, and its
length is called the Trotter number.
The $(d+1)$-dimensional classical system becomes
equivalent to the original $d$-dimensional quantum system when the
Trotter number is infinite.
As the Trotter number increases, the correlation length along the
Trotter direction increases, and the dynamics of the simulation slows down.

The simulation in the Q1D system is equivalent to the QMC simulation
if we regard the Trotter direction as the $c$ axis in the Q1D system.
For example,
the cluster-heat-bath algorithm in the Q1D system is equivalent to
the transfer-matrix MC method\cite{miya-tmmc,kagome} in QMC.
This is the main idea of this paper.
We know that the continuous imaginary-time loop flip algorithm of 
QMC\cite{qmc-loop1,qmc-loop2,qmc-looprev,totaqmc}
is very efficient.
Therefore, we apply this QMC algorithm to the Q1D simulation.
The correlated cluster along the $c$ axis is flipped by one update trial.
We do not suffer from the MC slowing-down due to the long correlation length.
The algorithm was successfully applied to the theoretical analysis on the
magneto-electric transitions in RbCoBr$_3$.\cite{totaRbCoBr3}
The numerical results quantitatively agree with the experimental results.
The estimates of the interaction parameters and proposals of new experiments
were made possible.

We consider the transverse-field Ising model in two dimension.
The Hamiltonian is written as
\begin{equation}
{\cal H}_q = 
-J\sum_{\langle j, k\rangle}
\sigma_j ^z \sigma_k^z - \Gamma \sum_j \sigma_j^x,
\end  {equation}
where 
$\sigma^x$ and $\sigma^z$ denote the Pauli spin operators,
$J$ denotes the exchange interaction parameter, 
and $\Gamma (>0)$ denotes the transverse field.
The indices, $j$ and $k$, denote the spin location on the two-dimensional
real-space plane throughout in this paper.
The bracket $\langle \cdots \rangle$ denotes the interacting spin pairs.
We apply the Suzuki-Trotter decomposition\cite{suzukitrotter} and
map the quantum system ${\cal H}_q$ to the effective classical system
${\cal H}_c$, which is written as
\begin{equation}
{\cal H}_c = 
\sum_{i=1}^m 
\left(
-\frac{J}{m}
\sum_{\langle j,k\rangle}
\sigma_{i,j} \sigma_{i,k}
-
\frac{\ln\coth(\frac{\beta\Gamma}{m})}{2\beta}
\sum_j
\sigma_{i,j} \sigma_{i+1,j}
\right)
.
\label{eq:mapHc}
\end  {equation}
Here, 
$m$ denotes the Trotter number, $\beta$ denotes the inverse temperature,
and $\sigma_{i,j}=\pm 1$ denotes the Ising spin.
The index, $i$, denotes the location along the Trotter direction throughout
in this paper.
The first term of this effective classical system denotes the 
exchange interaction between spins on the same Trotter slice.
The second term is the exchange interaction between spins at the same 
real-space site with the different (nearest-neighbor) Trotter slice.

The effective classical system can be regarded as the Q1D spin system:
\begin{equation}
{\cal H}_{\rm Q1D} = 
\sum_{i=1}^{L_c}
\left(
-J_{ab}
\sum_{\langle j,k\rangle}
\sigma_{i,j} \sigma_{i,k}
-J_c
\sum_j
\sigma_{i,j} \sigma_{i+1,j}
\right),
\label{eq:Q1D}
\end  {equation}
if we set
\begin{eqnarray}
J&=& mJ_{ab}, \label{eq:J}\\
\Gamma&=& \frac{m}{2\beta}\ln \coth[\beta J_c],\label{eq:G}\\
m &=& L_c.\label{eq:m}
\end  {eqnarray}
Here,
$i$ denotes the location along the $c$ axis,
$j$ and $k$ denote the location on the $ab$ plane, and
$L_c$ denotes the linear size along the $c$ axis.
The simulation in the Q1D system can be substituted for the QMC simulation with
$J, \Gamma$, and $m$ defined above.
The sign of $J_c$ is positive (ferromagnetic) in this expression.
In the case when it is negative (antiferromagnetic), we transform
it to the ferromagnetic one by changing the spin notation as
$\sigma_{i,j}\to (-1)^i\sigma_{i,j}$.

Let us consider the cluster algorithm of the Q1D system.
It is the interpretation of the QMC cluster algorithm \cite{totaqmc},
where the cluster is only defined along the Trotter direction.
We define a cluster using only the $J_c$ part of the Q1D
Hamiltonian, and consider the $J_{ab}$ part as the molecular field to
the cluster.
We may regard this algorithm as the Swendsen-Wang algorithm \cite{swendsen}
in one dimension [the $J_c$ part in Eq.~(\ref{eq:Q1D})]
under the molecular field [the $J_{ab}$ part in Eq.~(\ref{eq:Q1D})].
The ergodicity and the detailed-balance condition are guaranteed.

The updating procedure is as follows.
First, 
we select one location $j$ on the $ab$ plane, 
and consider the spins along the $c$ axis.
We define clusters by connecting the neighboring spins 
($\sigma_{i,j}$ and $\sigma_{i+1, j}$)
with the following probability $p_c$:
\begin{eqnarray}
p_c&=&
1-\exp[-2\beta J_c] 
~~~~(\sigma_{i,j}=\sigma_{i+1,j}), \label{eq:prob}\\
p_c&=&0  ~~~~~~~~~~~~~~~~~~~~~~~ (\sigma_{i,j}\ne\sigma_{i+1,j}).
\end  {eqnarray}
Let us number the cluster by $I$.
Second, we calculate the molecular field $h_I$
for each updating cluster $I$ as
\begin{equation}
h_I =   \sum _{i\in I}\sum_{\langle k\rangle} J_{ab}\sigma_{i,k},
\label{eq:hi}
\end  {equation}
where $i\in I$ denotes that $i$ belongs to the cluster $I$, 
and $\langle k \rangle$ denotes that $\sigma_{i,k}$ is interacting with
$\sigma_{i,j}$.
Finally,
we flip the cluster state with the following probability $p_I$:
\begin{equation}
p_I  = \frac{1} {\exp[2\beta\sigma_{i,j}h_I] + 1}.
\end  {equation}
We independently try this flip for each cluster.

The MC correlation time is reduced by this flip but the real
CPU time rather increases because we have to do the connecting procedures
for all spins along the $c$ axis.
We solve this problem by applying the continuous imaginary-time cluster
flip algorithm of QMC.\cite{qmc-loop2,qmc-looprev,totaqmc}
We neglect the discreteness of the spin location along the $c$ axis.
This approximation is possible when both cluster length and $L_c$ are very long.

\begin{figure}
\epsfxsize=7.5cm
\epsffile{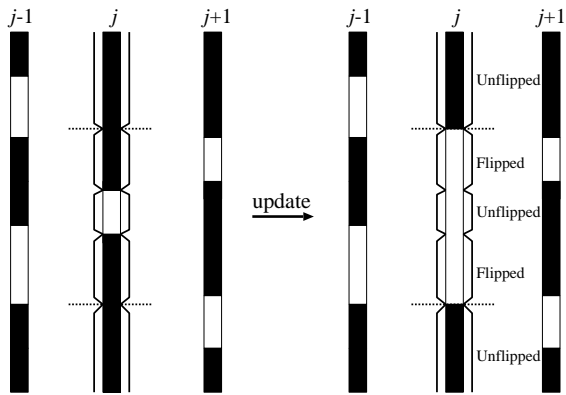}
\caption {
The updating procedure of the continuous $c$ axis version.
Black (white) rectangles depict the up-state (down-state) spin clusters.
The new-generated cluster edges are depicted by broken horizontal lines.
Brackets depict the clusters to be updated.
We update each cluster state independently using the probability $p_I$.
}
\label{fig:edge}
\end  {figure}

In the continuous version we focus on the locations of the cluster edges.
The probability that the spin pair of $\sigma_{i,j}=\sigma_{i+1,j}$ 
is not connected is
\begin{equation}
\exp[-2\beta J_c]
=\xi_c^{-1},
\end{equation}
where $\xi_c$ is regarded as the correlation length along the $c$ axis.
The average cluster size coincide with the correlation length $\xi_c$.
If we set 
$
L_c=L_{ab}\xi_c, 
$
the system roughly consists of $L_{ab}^3$ correlated clusters.
It is known that the cluster length obeys the Poisson 
distribution.\cite{qmc-looprev}
We generate the Poisson random numbers with the mean $\exp[2\beta J_c]$ and
regard them the cluster length.
Then, we place the cluster edges to the $c$ axis from bottom to top.
Combining these new-generated cluster edges and the already-existing ones, 
we apply the cluster flip with the probability $P_I$.
The procedure is shown in Fig.~\ref{fig:edge}

The present continuous $c$ axis version benefits from the memory
reduce and the CPU time reduce.
We do not need to memorize all the spin state.
Only the locations of the cluster edges and the spin state at the bottom 
edge are necessary.
The total memory use and the real CPU time are proportional to $L_{ab}^3$. 
Those for the single-spin-flip algorithm are proportional to
the total number of spins, $L_{ab}^3\xi_c$.
The efficiency gain, $\xi_c$,
becomes exponentially large at low temperatures.

We apply the continuous $c$ axis cluster flip algorithm to the 
stacked-triangular lattice antiferromagnetic Ising model.
It is a model system for the ABX$_3$-type compounds.%
\cite{CsCoBr3,CsCoCl3,visser,experi-mag,experi-f}
The Hamiltonian is written as follows.
\begin{eqnarray}
\mathcal{H}&=&
-2J_{\mathrm c}\sum_{i,j} S_{i,j}S_{(i+1),j}
-2J_{\mathrm 1}\sum_i\sum_{\langle jk\rangle}^{\rm n.n.}
S_{i,j}S_{i,k}
\nonumber \\
&&- 2J_{\mathrm 2}\sum_i\sum_{\langle jk\rangle}^{\rm n.n.n.}
S_{i,j}S_{i,k},
\end  {eqnarray}
where $S_{i,j}=\frac{1}{2}\sigma_{i,j}$ is the spin-1/2 Ising spins, and
$J_1 (J_2)$ denotes the nearest-neighbor (next-nearest-neighbor) exchange
interactions within the $ab$ plane.
We consider the case where both $J_c$ and $J_1$ are antiferromagnetic
($J_c, J_1 < 0)$, 
and $J_2$ is ferromagnetic ($J_2>0$).

It is known through the theoretical 
analyses\cite{shiba,matsubara-ina,koseki,todoroki}
that successive magnetic phase transitions occur.
The low-temperature magnetic structure is the ferrimagnetic state.
There exists a partially-disordered (PD) phase between the paramagnetic
phase and the ferrimagnetic phase.
In the PD phase, one of three sublattices is completely disordered, while
the other two sublattices take antiferromagnetic configurations.
It is considered that the phase transition between the paramagnetic phase 
and the PD phase is the second-order transition.
We refer to the transition temperature as $T_{\rm N1}$.

We compare the equilibration and the real CPU time of the present algorithm
with the results of the single-spin-flip algorithm.
We set $J_c=-97.4$K, $J_1=-2.44$K, $J_2=0.142$K, and perform the
simulation at $T=25$K. 
The system is in the ferrimagnetic phase at this temperature.
The linear lattice size of the $ab$ plane is set as $L_{ab}=95$.
The correlation length along the $c$ axis is roughly estimated as
$\xi_c\sim \exp[\beta |J_c|]= 49$, and the linear size along the
$c$ axis is set as $L_c=L_{ab}\xi_c=4655$.
The effective number of spins is more than 42 millions.
We observe the relaxation functions of the structure factors defined as
follows.
\begin{eqnarray}
f_{1/3}^2&=&\frac{1}{{8}}
\left\langle
\sum_{\eta=\alpha, \beta, \gamma} 
(m_{\eta} - m_{\eta+1})^2
\right\rangle,
\label{eq:f3s}
\\
f_{1}^2&=&\left\langle
(m_{\alpha}+m_{\beta}+m_{\gamma})^2
\right\rangle,
\label{eq:f1s}
\end  {eqnarray}
where $m_{\alpha}$, $m_{\beta}$, and $m_{\gamma}$ are three sublattice
magnetizations in the triangular lattice.
The 1/3-structure factor, $f_{1/3}^2$ 
takes a finite value when the ferrimagnetic state or the PD state is realized. 
It detects the phase transition between the PD phase and the paramagnetic phase.
The phase transition between the PD phase and the ferrimagnetic phase 
is detected by the structure factor, $f_1$.

\begin{figure}
\epsfxsize=7.5cm
\epsffile{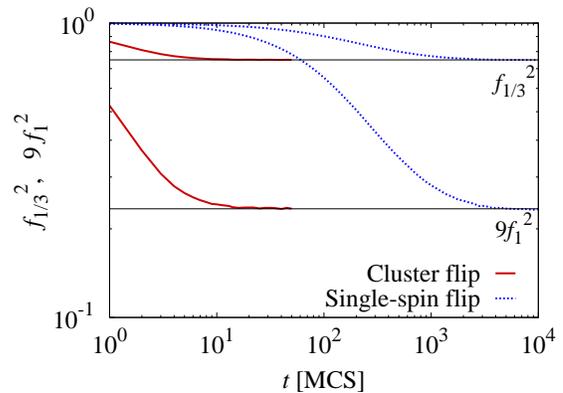}
\caption{(Color online)
The relaxation function of the structure factors, $f_{1/3}^2$ and
$9f_1^2$.
}
\label{fig:ner}
\end  {figure}

Figure \ref{fig:ner} shows the relaxation functions of both structure
factors.
We start the simulation from the perfect ferrimagnetic state, where
structure factors take $f_{1/3}^2=1$ and $f_1^2=1/9$.
The data of two algorithms converge to the same value.
It guarantees the equilibration of the simulation.
The cluster algorithm realizes the equilibrium state roughly 300 times
earlier than the single-spin-flip algorithm.
Table \ref{tab:cputime} compares the real CPU time.
The present cluster algorithm achieves the 15 times faster simulation.
This difference comes from the ratio $L_c/L_{ab}=\xi_c$.

\begin{table}
\begin{tabular}{c c c}
\hline
MC steps [MCS]~~~ &~~~ Cluster flip [s] ~~~& ~~~ Single-spin flip [s] ~~~\\
\hline
100 & 39   & 568   \\
200 & 77   & 1155   \\
500 & 195 & 2910 \\
10000 & 3919 & 58560 \\
\hline
\end {tabular}
\caption{
The real CPU time for each Monte Carlo step is compared.
The simulations were performed on the Core 2 Duo E6600 processor at 2.4GHz 
using the Intel compiler.
}
\label{tab:cputime}
\end  {table}

We focus on the transition temperature 
between the paramagnetic phase and the PD phase, $T_{\rm N1}$.
The chain-mean-field theory\cite{shiba} gives the relation among
$T_{\rm N1}$, $J_1$, and $J_c$, which is written as
\begin{equation}
1=\frac{\exp[\frac{|J_c|}{k_{\rm B}T_{\rm N1}}]}{2k_{\rm B}T_{\rm N1}}
(-3J_1+6J_2).
\label{eq:cmf}
\end  {equation}
Using the present cluster algorithm we estimate $T_{\rm N1}$ 
for various choices of $J_1/J_c$ 
ranging from 0.001 to 0.5 and $J_2/J_c$ ranging from -0.05 to -0.0015.
The behavior of $T_{\rm N1}$ with respect to $J_c, J_1$, and $J_2$ is obtained.
Here, the nonequilibrium relaxation method\cite{nerreview} is applied.
We obtain the transition temperature by the behavior of the relaxation
functions of the structure factor, $f_{1/3}^2$.
It converges to the finite value when the temperature is below $T_{\rm N1}$ 
and decays exponentially when the temperature is above $T_{\rm N1}$.
The algebraic decay is exhibited at $T_{\rm N1}$.

\begin{figure}
\epsfxsize=7.5cm
\epsffile{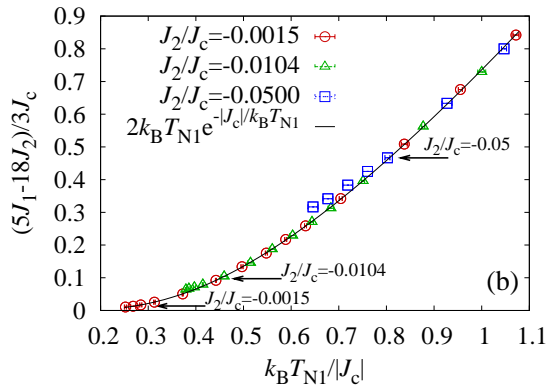}
\caption{
(Color online)
Relation between the exchange interaction parameters and the transition
temperature $T_{\rm N1}$ obtained by the Monte Carlo simulation.
Arrows depict the point where $|J_2|=|J_1|/2$ for each choice of $J_2/J_c$. 
The numerical results fall onto the single function as long as $|J_2|<|J_1|/2$.
}
\label{fig:tcpd}
\end  {figure}

We find that most of our numerical results are well-fitted by
the following expression.
\begin{equation}
1=\frac{\exp[\frac{|J_c|}{k_{\rm B}T_{\rm N1}}]}{2k_{\rm B}T_{\rm N1}}
\left(-\frac{5}{3}J_1+6J_2\right).
\label{eq:modify}
\end  {equation}
Only the coefficient of $J_1$ differs from the chain-mean-field result.
The change of the coefficient can be regarded as the reduction of the
effective coordination number.\cite{todo2006}
The fitting is plotted in Fig.~\ref{fig:tcpd}.
Arrows in the figure depicts the data when $|J_2|=|J_1|/2$
for each choice of $J_2$.
The data deviate from the relation, Eq.~(\ref{eq:modify}),
when $|J_2|>|J_1|/2$.
Since the chain-mean-field relation, Eq.~(\ref{eq:cmf}),
has been used to estimate the
interaction parameter from the experimental results, 
the present relation, Eq.~(\ref{eq:modify}), improves the estimate.

We have introduced the cluster flip algorithm suitable 
for the quasi-one-dimensional frustrated Ising spin systems.
The numerical efficiency is improved as we lower the temperature and/or
as we increase the anisotropy ratio, $|J_c/J_{ab}|$.
Other algorithms mostly fail in this situation.
The realistic simulations (or emulations) for real compounds are made possible.
The quantitative MC analyses to the experimental results may help developments
in the material science.
Simulations under the magnetic field is possible.
We may include the field term into the molecular field term, $h_I$.

\acknowledgments

The use of random number generator RNDTIK programmed by
Prof. N. Ito and Prof. Y. Kanada is gratefully acknowledged.

\end{document}